\documentclass[superscriptaddress]{revtex4}
\usepackage{graphicx}
\usepackage{color}

\begin{document}

\title{Enhanced nonlocal effects in metamaterials with moderate-index inclusions}

\author{Carlo Rizza}
\affiliation{Consiglio Nazionale delle Ricerche, CNR-SPIN, Via Vetoio 10, 67100 L'Aquila, Italy}

\author{Vincenzo Galdi}
\affiliation{Fields \& Waves Lab, Department of Engineering, University of Sannio, I-82100 Benevento, Italy}

\author{Alessandro Ciattoni}
\affiliation{Consiglio Nazionale delle Ricerche, CNR-SPIN, Via Vetoio 10, 67100 L'Aquila, Italy}

\begin{abstract}
We investigate a class of multilayered metamaterials characterized by moderate-index inclusions and low average permittivity. Via first-principle calculations, we show that in such scenario first- and second-order spatial dispersion effects may exhibit a dramatic and non-resonant enhancement, and may become comparable to the local response. Their interplay gives access to a wealth of dispersion regimes encompassing additional extraordinary waves and topological phase transitions. In particular, we identify a novel configuration featuring bound and disconnected isofrequency contours. Since they do not rely on high-index inclusions, our proposed metamaterials may constitute an attractive and technologically viable platform for engineering nonlocal effects in the optical range.
\end{abstract}

\maketitle

Metamaterials are artificial composites of dielectric and/or metallic inclusions in a host medium, which can be engineered so as to exhibit unconventional and/or tailored effective electromagnetic responses. As such, they constitute a unique platform for attaining novel optical mechanisms such as negative refraction, super- and hyper-lensing, invisibility cloaking, etc. \cite{Cai}. In the long-wavelength limit, the metamaterial optical response is typically described in terms of macroscopic phenomenological parameters, such as effective permittivity and permeability \cite{Smith}. However, if the inclusions are metallic and/or their size is not electrically small, nonlocal effects (i.e., spatial dispersion) may need to be accounted for, e.g., via the appearance of spatial derivatives of the fields in the effective constitutive relationships, or via wavevector-dependent constitutive parameters \cite{Landau}. The reader is referred to \cite{Silveirinha,Elser,Alu,Chebykin1,Chebykin2} for a representative sampling of nonlocal homogenization approaches available in the topical literature. It is worth stressing that two cornerstones in metamaterial science, namely, artificial electromagnetic chirality and optical magnetism, are manifestations of first- and second-order spatial dispersion, respectively \cite{Serd,Ciatt_1}. Artificial electromagnetic chirality (due to 3-D, 2-D and 1-D geometrical chirality \cite{Rizza_1}) yields interesting phenomena, such as giant optical activity, asymmetric transmission, and negative refractive index \cite{Li}. Artificial or optical magnetism may give rise to negative refractive index and backward-wave propagation \cite{Smith_2000}.

It is worth noting that second-order spatial dispersion is not fully equivalent to artificial magnetism \cite{Gorlach},
and this aspect has not been exploited to its fullest extent in applied science. Even though second-order and non-magnetic nonlocal effects are generally considered detrimental for several applications \cite{Taming}, they have been shown to support intriguing phenomena such as propagation of additional extraordinary waves \cite{Orlov1,Zayats} and topological transitions \cite{Gorlach,Orlov2,Mirmo}. Within this context, harnessing these nonlocal effects currently represents one of the grand challenges of metamaterial science.

Spatial dispersion is essentially ruled by the electrical size $\eta=\Lambda/\lambda$, with $\Lambda$ denoting a characteristic metamaterial scale (unit-cell size) and $\lambda$ the vacuum wavelength. Therefore, unlike conventional materials, metamaterials offer a natural way for enhancing nonlocality by tailoring the spatial distribution of the inclusions. There are several strategies to modify the hierarchy of spatial-dispersion orders: {\em i)} by optimizing shape and size of the inclusions \cite{Saenz}, {\em ii)} by exploiting plasmonic resonances \cite{Leon}, {\em iii)} by resorting to extreme regimes where permittivity and/or permeability are close to zero \cite{Rizza_1}, and {\em iv)} by embedding high-index dielectric inclusions, i.e., with relative permittivity $|\varepsilon_I| \sim 1/\eta^2$ \cite{Felbacq,Merlin}. The latter strategy has led to the development of low-loss, all-dielectric metamaterials with electric and magnetic responses \cite{Krasnok}.

Here, we propose a strategy to enhance both first- and second-order nonlocal effects, so that their interplay may become comparable to the local response, without resorting to plasmonic resonances. Such interplay is remarkably unusual in non-resonant scenarios,
since first- and second-order spatial-dispersion effects typically differ by orders of magnitude. Moreover, their symmetry properties are also different, since first-order nonlocality is ruled by geometric chirality, whereas second-order nonlocality is not.

We show that the above strategy can be implemented in multilayered metamaterials characterized by {\em moderate-index} inclusions (MII), i.e., with relative permittivity $|\varepsilon_I| \sim 1/\eta$, in conjunction with a relatively low average relative permittivity ($\ll 1/\eta$). Via first-principle derivations, we show that transverse magnetic (TM) waves are ruled by a biquadratic, highly tunable dispersion relationship that yields a broad variety of phenomena, encompassing additional extraordinary waves and topological phase transitions. Besides the well-known hyperbolic, elliptic and mixed regimes, we identify a topologically nontrivial dispersion regime characterized by bounded and multiply-connected isofrequency contours accompanied by additional extraordinary waves. The proposed metamaterials, which combine a fairly simple geometrical structure with relaxed (moderate-index) requirements on the material constituents, may provide a versatile and technologically viable platform for dispersion engineering, especially attractive in spectral regions (e.g., optical range) where extremely high-index materials are not available.

\begin{figure}
\centering
\includegraphics[width=0.45\textwidth]{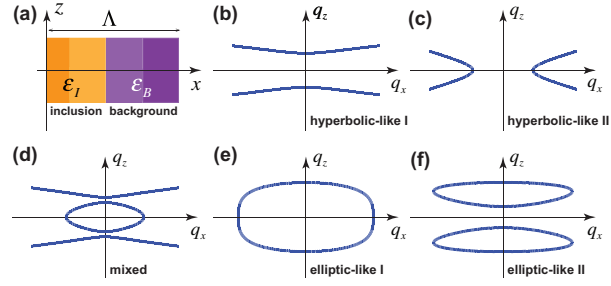}
\caption{(a) Schematic of the metamaterial unit cell. (b)--(f) Representative isofrequency contours attainable from Eq. (\ref{disp2}).} \label{Figure1}
\end{figure}

With reference to the schematic in Fig. \ref{Figure1}(a), we consider a 1-D multilayered metamaterial, periodically stratified along the $x$-axis, with a unit cell of thickness $\Lambda$ consisting of two (possibly composite) regions that we generically indicate as ``background'' and ``inclusion''. All fields and quantities are assumed as independent of $y$, and the metamaterial is described by a piecewise continuous relative permittivity distribution $\varepsilon(x)$, denoted by $\varepsilon_B$ and $\varepsilon_I$ within the background and the inclusion regions, respectively. In the presence of a monochromatic electromagnetic field (with suppressed $e^{-i \omega t}$ time-dependence) whose vacuum wavelength $\lambda$ is much larger than the unit-cell period, the small ratio $\eta = \Lambda / \lambda$ rules the homogenization of the metamaterial response. As anticipated, we are interested in the MII regime characterized by the basic requirements:
\begin{eqnarray}
\label{requirements}
\label{req1}
|\varepsilon_B| &\ll& 1/\eta, \quad |\varepsilon_I| \sim 1/\eta,\\
\label{req2}
|\overline{\varepsilon(x)}| &\ll& 1/ \eta,
\end{eqnarray}

with the overline indicating the average over the unit cell. As detailed in \href{link}{Supplement 1},
a rigorous, multi-scale analysis of TM-wave ($y$-directed magnetic field) propagation, in the long-wavelength limit $\eta \ll 1$ and the MII regime, shows that the leading-order metamaterial response is governed by the effective constitutive relationships
\begin{eqnarray}
\label{response}
\overline{D}_x &=& \epsilon_0 \left( \epsilon _{||} \overline{E}_x +  \frac{\kappa}{k_0} \frac{\partial \overline{E}_z}{\partial z}   \right), \\
\overline{D}_z &=& \epsilon_0 \left( \epsilon _{\perp} \overline{E}_z -  \frac{\kappa }{k_0} \frac{\partial \overline{E}_x}{\partial z} + \frac{\delta}{k_0^2} \frac{\partial^2 \overline{E}_z}{\partial z^2} \right), \\
\overline{B}_y &=& \mu_0 \overline{H}_y,
\end{eqnarray}
with $\epsilon_0$, $\mu_0$ and $k_0=\omega\sqrt{\epsilon_0\mu_0}=2\pi/\lambda$ denoting the vacuum permittivity, permeability and wavenumber, respectively, and
\begin{eqnarray}
\label{par_eff1}
\epsilon_{||} &=& \left[ \overline{ \varepsilon^{-1}(x)   } \right]^{-1}, \:
\epsilon_{\perp} = \overline{\varepsilon(x)} +  \sum_n \frac{A_{-n} A_{n}}{n^2},\\
\label{par_eff2}
\kappa &=& i \epsilon _{||} \sum_n \frac{A_{-n} B_{n}}{n}, \\
\label{par_eff3}
\delta &=& \frac{1}{\epsilon_{||}} \left( \sum_n  \frac{A_{-n} A_{n}}{n^2} - \kappa^2 \right) +  \sum_{n,m} \frac{A_{m} A_{-(m+n)} B_{n}}{m(m+n)},
\end{eqnarray}
where the summations are intended over all integers $n,m \neq 0$, with $n \neq -m$, and $A_n / \eta$ and $B_n$ are the Fourier coefficients of the functions $\varepsilon(x)$ and $\varepsilon^{-1}(x)$, respectively (with $|A_n|,\: |B_n| \ll 1/\eta$). Equations (\ref{response}) and (\ref{par_eff}) summarize the main result of this study, and they show that the MII regime, in addition to the local response, exhibits both first- and second-order spatial dispersion. By using the Serdyukov-Fedorov transformation method \cite{Serd}, it can be shown that the obtained first-order spatial dispersion is equivalent to 1-D electromagnetic chirality \cite{Rizza_1}, and that the second-order spatial dispersion is non-magnetic (i.e., there is no transformation yielding a non-vanishing magnetic permeability). It is crucial to stress that Eqs. (\ref{response}) {\em do not} contain the small parameter $\eta$. From the physical viewpoint, this yields an unprecedented regime where local response and nonlocal effects may interplay at comparable levels. By contrast, in media with shallow dielectric modulation, first- and second-order spatial-dispersion contributions scale as $\sim \eta$ and $\sim \eta^2$, respectively, and hence are too weak to appear in the leading-order term of the effective response \cite{Merlin}.

Our proposed MII regime in Eqs. (\ref{requirements}) forbids the onset of resonances that would produce a dominant local response, and, at the same time, it triggers an anomalous enhancement of nonlocal effects which strengthens (in different ways) the relative weight of standard first- and second-order spatial-dispersion contributions, while leaving higher-order spatial-dispersion contributions at a negligibly weak level. To gain some physical insight into the underlying mechanisms, we start analyzing the local response. Due to the requirements in Eqs. (\ref{req1}), the average of $ \varepsilon^{-1}(x)$ is dominated by the background contribution $\varepsilon_B$, so that $|\epsilon_{||} | \ll 1/\eta$ [see Eq. (\ref{par_eff1})]. This is a key aspect characterizing the MII regime since, in the presence of a uniformly large dielectric spatial modulation, $|\epsilon_{||}|$ can be exceedingly large so as to yield superlensing effects \cite{Rizza_2} which are unaffected by spatial dispersion. The requirements in  Eqs. (\ref{req1}), attainable by suitably tailoring the inclusion dielectric profile (see our discussion and examples below), imply that the first contribution $\overline{\varepsilon(x)}$ to $\epsilon_{\perp}$ in Eq. (\ref{par_eff1}) is not large. Moreover, the Fourier coefficients of $\varepsilon(x)$ (with $n \neq 0$) are $A_n / \eta$ with $|A_n| \ll 1/\eta$, and hence $|\epsilon_{\perp}| \ll 1/\eta$. In other words, even if the dielectric modulation is large, the MII regime is characterized by local effective permittivities that are sufficiently small so as not to overshadow the spatial-dispersion effects. Considering now the nonlocal response,
the requirements in Eqs. (\ref{req1}) imply that all Fourier coefficients of $\varepsilon^{-1}(x)$ are dominated by the background permittivity, and hence $|B_n| \ll 1/\eta$. As a consequence, from Eqs. (\ref{par_eff2}) and (\ref{par_eff3}), the strength of the nonlocal response may become comparable to the local one (see also \href{link}{Supplement 1} for a heuristic explanation of the enhancement effects).
It is also worth emphasizing that, although the MII requirements in Eqs. (\ref{requirements}) do not explicitly provide indications on the permittivity signs, the most significant spatial-dispersion enhancements [i.e., respectable values of $\kappa$ and $\delta$ in Eqs. (\ref{par_eff})] are obtained in configurations mixing positive- and negative-permittivity constituents. From the physical viewpoint, this is not surprising, since these configurations may support surface-plasmon-polaritons propagating at the layer interfaces, whose coupling is known to yield strong nonlocal effects \cite{Kidwai}.


In order to illustrate the physical implications of the strong MII nonlocal response, we analyze the dispersion relationship of TM waves $H_y=U_y e^{ik_0(q_x x+ q_z z)}$, which readily follows from the constitutive relationships in Eqs. (\ref{response}), viz.,
\begin{equation}
\label{disp2}
\delta q_z^4 -\left( \epsilon _{\perp} + \kappa^2 +  \delta  \epsilon _{||} \right) q_z^2 - \epsilon _{||} q_x^2 + \epsilon _{||} \epsilon _{\perp} = 0.
\end{equation}
A first important consequence of second-order spatial-dispersion (represented by $\delta$) is that such equation is {\em biquadratic}. Moreover, the four coefficients in Eq. (\ref{disp2}) depend on the four independent effective parameters in Eqs. (\ref{par_eff}) which, in turn, may assume comparable values. This implies that, in principle, every possible dispersion regime arising from Eq. (\ref{disp2}) is accessible.
\begin{figure}
\centering
\includegraphics[width=0.45\textwidth]{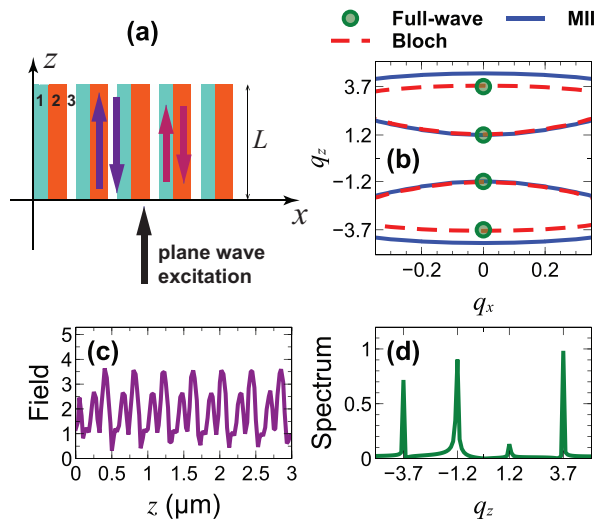}
\caption{First example (details in the text). (a) Schematic of the MII metamaterial slab, and excitation of additional extraordinary waves (thick arrows). (b) Comparison among the isofrequency computed via the MII effective model (blue-solid curves),  Bloch theory (red-dashed curves), and full-wave simulations (only for $q_x=0$, green circle markers). (c) Magnitude distribution of the average magnetic field $|\overline{H}_y(z)|$, normalized by the incident field. (d) Corresponding spatial spectrum (magnitude), in arbitrary units.} \label{Figure2}
\end{figure}
Figures \ref{Figure1}(b)--\ref{Figure1}(f) qualitatively illustrate, in terms of isofrequency contours, the five possible and topologically different dispersion regimes that can be attained: hyperbolic-like I and II, mixed, elliptic-like I and II, respectively. It is remarkable that the relatively simple class of MII metamaterials is able to support such a variety of dispersion regimes, comprising both bounded [Figs.  \ref{Figure1}(e) and \ref{Figure1}(f)] and unbounded [Figs. \ref{Figure1}(b), \ref{Figure1}(c) and \ref{Figure1}(d)] propagating wave spectra whose isofrequency contours can be either connected [Fig. \ref{Figure1}(e)] or disconnected [Figs. \ref{Figure1}(b), \ref{Figure1}(c), \ref{Figure1}(d) and \ref{Figure1}(f)].
Elliptic-hyperbolic mixed regimes such as in Fig. \ref{Figure1}(d)
 have been predicted and experimentally observed \cite{Zayats,Gorlach,Mirmo}, together with additional-wave manifestations, in different metamaterial configurations where second-order spatial dispersion plays a key role in the absence of electromagnetic chirality.
To the best of our knowledge, the elliptic-like II regime in Fig. \ref{Figure1}(f) has never been predicted in 1-D metamaterials; its distinctive feature is
the presence of two propagating waves with the same transverse wavenumber (i.e., an additional extraordinary wave) within a limited $q_x$ spectral region.


To validate and calibrate the MII effective model in Eqs. (\ref{response}), we compare its predictions with rigorous Bloch theory \cite{Joannopoulos} and full-wave numerical simulations \cite{Comsol} (see \href{link}{Supplement 1} for details). As a first illustrative example, we consider a unit cell comprising three homogeneous layers, two of which (labeled as $1$ and $2$) constituting the inclusion, and the third one (labeled as $3$) representing the background [see Fig. \ref{Figure2}(a)]. We consider a parameter configuration with $\eta=0.1$, layer relative permittivities $\epsilon_1=3.8$, $\epsilon_2=-4.2$ and $\epsilon_3=-0.2$, and filling fractions $f_1=f_2=0.4$ and $f_3 = 0.2$. Such a design fulfills the MII requirements in Eqs. (\ref{requirements}), since $|\varepsilon_B| = 0.2 \ll 1/\eta$, $|\varepsilon_I| \simeq 4 \sim 1/\eta$ and $| \overline{\varepsilon(x)}| = |f_1 \epsilon_1 + f_2 \epsilon_2 + f_3 \epsilon_3| = 0.2 \ll 1/\eta$. For the chosen parameter configuration, we obtain from Eqs. (\ref{par_eff}) $\epsilon_{\perp}=-9.2 \cdot 10^{-2}$, $\epsilon_{||}=-1$, $\kappa=-0.41$ and $\delta=3.4 \cdot 10^{-3}$. This
yields the elliptic-like II dispersion regime in Fig. \ref{Figure1}(f), where additional waves are expected, as sketched in Fig. \ref{Figure2}(a).
Figure \ref{Figure2}(b) compares the isofrequency contours computed from Eqs. (\ref{disp2}) with the Bloch-theory \cite{Joannopoulos} solution. A fairly good agreement is evident, as well as the occurrence of an additional extraordinary wave, which is accurately captured by the MII effective model. Note also that $\overline{\varepsilon(x)} = -0.2$ and $\left[ \overline{\varepsilon^{-1}(x)} \right]^{-1} = -1$, so that the standard (local) effective-medium theory (EMT) \cite{Cai} predicts that plane-wave propagation is forbidden (evanescent) in the medium, in stark contrast with the above evidence.
It is worth stressing that, in contrast to high-index-inclusion metamaterials \cite{Felbacq}, such regime dominated by spatial dispersion is accessed here by means of {\em moderate} values of the layer permittivities. As a further validation check, we also analyzed via full-wave simulations a MII metamaterial slab of thickness $L=10$ $\mu$m, with the above parameters and period $\Lambda=0.1$ $\mu$m, under normally incident (along $z$) plane-wave illumination with wavelength $\lambda=1$ $\mu$m [see Fig. \ref{Figure2}(a)]. Figure \ref{Figure2}(c) shows the magnitude of the average field distribution $|\overline{H}_y(z)|$, whose standing-wave pattern reveals the occurrence of different spatial harmonics. This is more evident in the corresponding spatial spectrum (magnitude) shown in Fig. \ref{Figure2}(d), which displays four distinct peaks corresponding to two forward and two backward propagating waves, thereby proving the actual excitability of additional extraordinary waves in a finite-thickness slab. For a more quantitative assessment,
the normalized wavenumbers $q_z$ corresponding to the spectral peaks are superimposed (circle markers) on the isofrequency contours in Fig. \ref{Figure2}(b), showing a good agreement with those predicted by the above description of an unbounded MII metamaterial.


\begin{figure}
\centering
\includegraphics[width=0.45\textwidth]{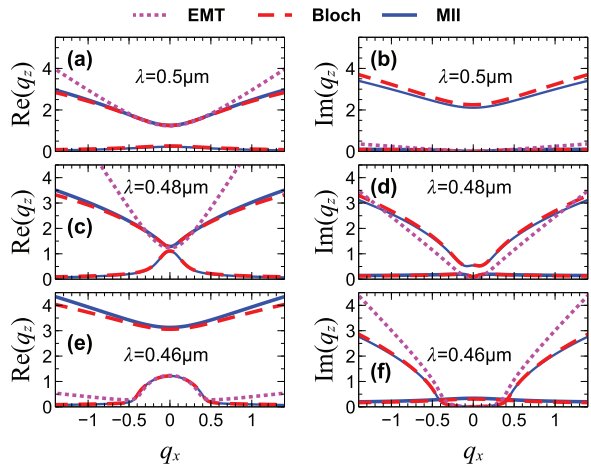}
\caption{Second example (details in the text). Complex isofrequency contours at three representative wavelengths (indicated in the panels), computed via the MII effective model (blue-solid curves), Bloch theory (red-dashed curves), and EMT (magenta-dotted curves). Due to symmetry, only the branches pertaining to forward-propagating waves are shown.} \label{Figure3}
\end{figure}

The existence of various different dispersion regimes implies that MII metamaterials may also host unusual topological transitions. As a second representative example, we consider a three-layer unit-cell of period $\Lambda=0.05$ $\mu$m, with material constituents chosen as silver (labeled as 1, and playing the role of the inclusion),  silicon dioxide (SiO$_2$) and air (labeled as 2,3 respectively, and playing the role of the background), with filling fractions $f_1 = 0.15$, $f_2 = 0.33$, and $f_3 = 0.52$. For silver, we utilize the Drude model $\epsilon_1=\epsilon_b-\omega_p^2/(\omega^2+i\alpha \omega)$ (with $\epsilon_b=5.26$, $\alpha=7.06 \cdot 10^{13}$ s$^{-1}$, $\omega_p=1.45 \cdot 10^{16}$ s$^{-1}$ \cite{JC}), whereas for SiO$_2$ ($\epsilon_2$) we employ an empirical Sellmeier equation \cite{Malitson}.
Figure \ref{Figure3} shows the complex isofrequency contours (real and imaginary parts of $q_z$ as a function of $q_x$), at three representative wavelengths $\lambda= 0.5, 0.48, 0.46 \: \mu m$ for which the MII requirements in Eqs. (\ref{requirements}) are fulfilled. Once again, the agreement between the MII effective model and Bloch theory is remarkably good at any wavelength, whereas the EMT provides an inadequate description, as it captures only a portion around $q_x=0$ of a single connected component of the isofrequency contours. Note that at, $\lambda = 0.5 \: \mu m$, the dispersion is of  hyperbolic-like I type, since the additional branches [lower in Fig. \ref{Figure3}(a), and upper in Fig. \ref{Figure3}(b)] are generated by the presence of loss (they disappear in the lossless limit). On the other hand, at $\lambda = 0.46 \: \mu m$ [Figs. \ref{Figure3}(e) and \ref{Figure3}(f)], a mixed regime is observed, and this indicates that a topological transition between the two regimes occurs. As also observed in Ref. \cite{Orlov2}, such transition is accompanied by an intermediate state at $\lambda = 0.48 \: \mu m$ [see Figs. \ref{Figure3}(c) and \ref{Figure3}(d)] where the two dispersion branches of the mixed regime merge at $q_x=0$, thereby yielding a wavevector degeneracy and an associated Dirac point.

%
In conclusion, we have introduced a class of 1-D MII metamaterials that may exhibit an anomalous electrodynamic behavior driven by nonlocal effects. By comparison with other strategies for spatial-dispersion enhancement, our proposed framework relaxes the requirements on the refractive-index of the inclusions, by affording moderate values. In essence, the specific structure of the MII requirements in Eqs. (\ref{requirements}) crucially alters the hierarchy of spatial-dispersion orders, leading to a previously unexplored interplay between first- and second-order nonlocal effects, whose strength may become comparable to the local response. This in turns leads to a variety of different dispersion regimes supporting exotic phenomena such as topological phase transitions and additional extraordinary waves. In view of their technological viability and potential versatility, we believe that MII multilayered metamaterials, and their possible higher-dimensional extensions, may play a pivotal role for conceiving novel platforms for dispersion-engineering applications over a broad electromagnetic spectrum, including the optical range.

\section*{Supplementary Material}
\subsection{Multiscale Homogenization Theory}
Referring to the schematic in Fig. 1(a), we consider a 1-D multilayered metamaterial with a unit-cell of spatial period $\Lambda$, described by a periodic relative permittivity distribution $\varepsilon(x)=\varepsilon(x+\Lambda)$. Moreover, we assume monochromatic, transverse-magnetic (TM) plane-wave propagation, with relevant field components $E_x(x,z)$, $E_z(x,z)$, $H_y(x,z)$ satisfying the source-free Maxwell's equations
\begin{eqnarray}
\label{Max}
\partial_z E_x - \partial_x E_z &=& i \omega B_y, \\
\partial_z H_y &=& i \omega D_x, \\
\partial_x H_y & = & - i \omega D_z,
\end{eqnarray}
where
\begin{eqnarray}
\label{cost}
D_x&=&\epsilon_0 \varepsilon E_x, \quad
D_z=\epsilon_0 \varepsilon E_z, \\
B_y&=&\mu_0 H_y,
\end{eqnarray}
and all symbols are already defined in the main text.  To investigate the long-wavelength limit $\Lambda\ll\lambda$,  and to obtain an effective medium description, we introduce the dimensionless small parameter $\eta=\Lambda/\lambda\ll 1$ and the fast-scale coordinate $X=x/\eta$, and we subsequently treat this coordinate as independent from the slow-scale variables $x$ and $z$. As anticipated in the main text, we consider a moderate-index inclusion (MII) parameter configuration fulfilling the requirements in Eqs. (1). In particular, the requirements in Eqs. (1a) imply that $|\varepsilon^{-1}(x)| \ll 1/\eta$ and $\varepsilon-\overline{\varepsilon(x)} \sim 1/\eta$. Accordingly, we address the homogenization of a specific profile $\epsilon(X)=\varepsilon(x)$ (depending only on the fast-scale coordinate $X$) and characterized by the properties
\begin{eqnarray}
\label{Kap}
\epsilon &=&  \overline{\epsilon} + \frac{1}{\eta} \widetilde{\epsilon}(X), \\
\epsilon^{-1} &=&  \overline{ {\epsilon}^{-1}} + \widetilde{\epsilon^{-1}}(X),
\end{eqnarray}
where the tilde indicates the rapidly varying zero-mean residual. In order to prevent the occurrence of secular terms in the limit $\eta \rightarrow 0$, we consider
\begin{eqnarray}
\label{Ans}
E_x&=&\overline{E}_x(x,z)+\widetilde{E}_x(x,z,X), \\
E_z&=&\overline{E}_z(x,z)+\eta \widetilde{E}_z(x,z,X),\\
H_y&=&\overline{H}_y(x,z)+\widetilde{H}_y(x,z,X),
\end{eqnarray}
where $\overline{A}$ and $\tilde{A}$ are the slowly and rapidly varying contributions to the field component $A$, respectively.  This constitutes the basic Ansatz of the two-scale homogenization method detailed hereafter, where each component $A$ depends on the fast-scale and slow-scale variables, and each electromagnetic field component in Eqs. (\ref{Ans}) is a suitable Taylor expansion up to the first order in the parameter $\eta$.
By substituting Eqs. (\ref{Ans}) and (\ref{Kap}) into Eqs. (\ref{Max}), and separating the fast and slow contributions, we obtain
\begin{eqnarray}
\label{slow}
&& \partial_z \overline{E}_x - \partial_x \overline{E}_z = i \mu_0 \omega \overline{H}_y, \\
&& \overline{\epsilon^{-1}}  \partial_z \overline{H}_y + \overline{ \widetilde{\epsilon^{-1}} \partial_z \widetilde {H}_y  } = i \omega \epsilon_0  \overline{E}_x, \\
&& \partial_x \overline{H}_y = -i \omega \epsilon_0 \left( \overline{\epsilon} \overline{E}_z + \overline{ \widetilde{\epsilon}  \widetilde {E}_z }  \right),
\end{eqnarray}
and
\begin{eqnarray}
\label{rapid}
&& \partial_z \widetilde{E}_x - \partial_x \widetilde{E}_z - i \mu_0 \omega \widetilde{H}_y = 0, \\
&& i \omega \epsilon_0  \widetilde{E}_x - \epsilon^{-1} \partial_z \widetilde {H}_y + \overline{  \widetilde{ \epsilon^{-1}} \partial_z  \widetilde{H}_y} = \widetilde{\epsilon^{-1}}  \partial_z \overline{H}_y, \\
&& \partial_X \widetilde{H}_y = -i \omega \epsilon_0 \widetilde{\epsilon} \overline{E}_z.
\end{eqnarray}
By introducing the Fourier series
\begin{equation}
\widetilde{\epsilon}=\sum_{{n=-\infty \\ n \neq 0}}^{+\infty} A_n e^{i n k_0 X}, \quad
\widetilde{\epsilon^{-1}}=\sum_{{n=-\infty \\ n \neq 0}}^{+\infty} B_n e^{i n k_0 X},
\end{equation}
we solve Eqs. (\ref{rapid}), and thereby obtain the rapidly varying contributions in terms of the slow (average) components.
By substituting the expressions of the rapidly varying contributions $\widetilde{E}_x$, $\widetilde{E}_z$ and $\widetilde{H}_y$ [derived by Eqs. (\ref{rapid})] into Eqs. (\ref{slow}), we then obtain the equations describing the dynamics of the average fields in the long-wavelength limit. These equations can be rearranged in the form of Maxwell's equations
\begin{eqnarray}
\label{Max_eff}
\partial_z \overline{E}_x - \partial_x \overline{E}_z &=& i \omega \overline{B}_y, \\
\label{Max_eff2}
\partial_z \overline{H}_y &=& i \omega \overline{D}_x, \\
\partial_x \overline{H}_y & = & - i \omega \overline{D}_z,
\end{eqnarray}
where the effective constitutive relationships are given by
\begin{eqnarray}
\label{cost_eff_1}
\overline{D}_x &=& \epsilon_0 \left( \epsilon _{||} \overline{E}_x + \frac{\kappa }{k_0} \partial_z \overline{E}_z  \right), \\
\overline{D}_z &=& \epsilon_0 \left( \epsilon _{\perp} \overline{E}_z + i \frac{\kappa Z_0}{ \epsilon _{||}k_0^2} \partial_z^2 \overline{H}_y  + \frac{\beta}{\epsilon _{||} k_0^2} \partial_z^2 \overline{E}_z \right), \\
\overline{B}_y &=& \mu_0 \overline{H}_y.
\end{eqnarray}
In Eqs. (\ref{cost_eff_1}), $Z_0=\sqrt{\mu_0/\epsilon_0}$ denotes the vacuum characteristic impedance, and we have introduced the effective constitutive parameters
\begin{eqnarray}
\label{par_eff}
\epsilon _{||} \!\!\!\!&=&\!\!\!\! \left(\overline{ \epsilon^{-1}}\right)^{-1},
\quad \epsilon _{\perp} =  \overline{\epsilon}+  \sum_{{n=-\infty, n \neq 0}}^{+\infty} \frac{A_{-n} A_{n}}{n^2}, \\
\kappa \!\!\!\!&=&\!\!\!\! i \epsilon _{||} \sum_{{n=-\infty, n \neq 0}}^{+\infty} \frac{A_{-n} B_{n}}{n}, \\
\beta \!\!\!\!&=&\!\!\!\! \sum_{{n=-\infty, n \neq 0}}^{+\infty} \Bigg [ \frac{A_{-n} A_{n}}{n^2} +  \epsilon _{||} \sum_{{m=-\infty, m \neq 0, -n}}^{+\infty}  \frac{A_{m} A_{-n-m} B_{n}}{m(m+n)} \Bigg ].
\end{eqnarray}
By using Eq. (\ref{Max_eff2}), the constitutive relations in Eqs. (\ref{cost_eff_1}) can be recast in the standard Landau-Lifshitz form \cite{Landau}, viz.,
\begin{eqnarray}
\label{cost_eff_2}
\overline{D}_x &=& \epsilon_0 \Bigg( \epsilon _{||} \overline{E}_x +  \frac{\kappa}{k_0} \partial_z \overline{E}_z  \Bigg), \\
\overline{D}_z &=& \epsilon_0 \Bigg( \epsilon _{\perp} \overline{E}_z -  \frac{\kappa }{k_0} \partial_z \overline{E}_x + \frac{\delta}{k_0^2} \partial_z^2 \overline{E}_z \Bigg), \\
\overline{B}_y &=& \mu_0 \overline{H}_y,
\end{eqnarray}
where
\begin{equation}
\label{delta}
\delta=\frac{\beta-\kappa^2}{\epsilon_{||}}.
\end{equation}
Equations (\ref{cost_eff_2}) are those reported as Eqs. (2) in the main text. As shown by Eqs. (\ref{cost_eff_2}), the electromagnetic response is affected by first- and the second-order spatial dispersion, i.e., the first- and second-order derivatives of the electric field describe the medium nonlocality. By using the Serdyukov-Fedorov transformation method \cite{Serd}, it can be shown that
the first-order spatial-dispersion contribution is equivalent to a reciprocal magneto-electric coupling (extrinsic chiral response) whereas the second-order spatial dispersion contribution cannot be cast as a magnetic term.
It is important to emphasize that
first- and second-order spatial dispersion terms in Eqs. (\ref{cost_eff_2}) are not weighted by the small parameter $\eta$, which indicates an anomalous enhancement.

\subsection{Heuristic Interpretation of the Anomalous Spatial-Dispersion Enhancement}
The anomalous enhancement of first- and second-order spatial-dispersion effects can be heuristically grasped from the effective medium response of a multilayered metamaterial with shallow dielectric modulation. Following the approach reported in Ref. \cite{Ciatt_1}, we obtain the effective constitutive relationships for TM waves propagating in a multilayered metamaterial, viz.,
\begin{eqnarray}
\label{standard}
\overline{D}_x &=& \epsilon_0 \left( \epsilon _{||}^S \overline{E}_x + \frac{\kappa^S }{k_0} \partial_z \overline{E}_z  +   \frac{\delta^S_a}{k_0^2} \partial^2_{x,z} \overline{E}_z + \frac{\delta^S_b}{k_0^2} \partial^2_z \overline{E}_x \right), \quad \quad \\
\overline{D}_z &=& \epsilon_0 \left( \epsilon^S_{\perp} \overline{E}_z - \frac{\kappa^S }{k_0} \partial_z \overline{E}_x
+  \frac{\delta^S_a}{k_0^2} \partial^2_{x,z} \overline{E}_x + \frac{\delta^S}{k_0^2}
\partial^2_z \overline{E}_z \right), \quad \quad \\
\overline{B}_y &=& \mu_0 \overline{H}_y.
\end{eqnarray}
The constitutive relationships in the above scenario resemble those in Eqs. (2) [and Eqs. (\ref{cost_eff_2})], but without some second-order spatial-dispersion contributions (i.e., $\delta_a=0$ and $\delta_b=0$), and with parameters (identified by the superscript $^S$)
\begin{eqnarray}
\epsilon_{||}^S &=& \left[ \overline{ \varepsilon^{-1}} \right]^{-1},
\quad \epsilon_{\perp}^S = \overline{\varepsilon}+\eta^2 \sum_{ {n=-\infty, n\neq 0}}^{\infty} \frac{a_{-n} a_n}{n^2},\\
\label{kappaS}
\kappa^S \!\!\!\!&=&\!\!\!\! i \eta \epsilon _{||} \sum_{ {n=-\infty, n\neq 0}}^{\infty} \frac{a_{-n} b_{n}}{n}, \\
\label{betaS}
\beta^S \!\!\!\!&=&\!\!\!\! \sum_{ {n=-\infty, n \neq 0}}^{+\infty} \Bigg [ \frac{a_{-n} a_{n}}{n^2} +  \epsilon _{||} \sum_{ {m=-\infty, m \neq 0, -n}}^{+\infty}  \frac{a_{m} a_{-n-m} b_{n}}{m(m+n)} \Bigg ], \\
\label{deltaA}
\delta^S_a \!\!\!\!&=&\!\!\!\! \eta^2 \epsilon _{||} \sum_{ {n=-\infty, n\neq 0}}^{\infty} \frac{a_{-n} b_{n}}{n^2}, \\
\label{deltaB}
\delta^S_b \!\!\!\!&=&\!\!\!\! - \eta^2 \left(\epsilon_{||}^S\right)^2 \sum_{ {n=-\infty, n \neq 0}}^{+\infty}   \sum_{ {m=-\infty, m \neq 0}}^{+\infty} \frac{a_{n-m} b_{-n} b_{m}}{n m},\\
\label{deltaS}
\delta^S\!\!\!\!&=&\!\!\!\!\frac{\beta^S-\left(\kappa^S\right)^2}{\epsilon^S_{||}},
\end{eqnarray}
where $a_n$, $b_n$ are the Fourier coefficients of $\varepsilon(x)$ and $\varepsilon^{-1}(x)$, respectively. For this configuration, spatial dispersion is {\em weak} since $|\kappa^S| \sim \eta\ll 1$, and $|\delta^S| \sim |\delta^S_a| \sim |\delta^S_b| \sim \eta^2 \ll 1$ in view of the low dielectric contrast. The crucial observation to make is that if, in Eq. (\ref{kappaS}) we formally replace $a_n$ with $A_n / \eta$ and $b_n$ with $B_n$, we fully recover the expression of $\kappa$ in Eq. (3a). This provides an insightful interpretation of the MII mechanism, suggesting that it {\em effectively elevates} the first-order spatial-dispersion effects from the $\eta$ order to the leading one. This observation also sheds light on the enhancement of second-order spatial-dispersion which, in a low-dielectric-contrast medium, is very weak ($\sim \eta^2$) with contributions containing either two $a_n$ or two $b_n$ terms. The above illustrated mechanism of elevating to the leading order the terms $a_n = A_n / \eta$ (for $n \neq 0$) likewise elevates the contributions containing two $a_n$ terms, and this explains their appearance in Eqs. (3). On the other hand, the terms proportional to $\delta^S_a$ and $\delta^S_b$ do not appear in Eqs. (3) since they contain a single $a_n$ term. Moreover, it is interesting to note that the terms proportional to $\delta^S_a$ correspond to the off-diagonal components of the wavevector-dependent effective permittivity in the approach proposed in Refs. \cite{Chebykin1,Chebykin2}, whereas the term proportional to $\delta^S_b$ is equivalent to a magnetic contribution and it can be strongly enhanced in the plasmonic-resonance condition ($|\epsilon_{||}|\gg1$) \cite{Pendry_00}.

\subsection{Details on Numerical Simulations}
In the results shown in Figs. 2 and 3, the rigorous Bloch-theory-based electromagnetic solution is implemented via a standard transfer-matrix approach \cite{Born}, by solving an eigenvalue problem with phase-shift boundary conditions \cite{Joannopoulos}.

The full-wave numerical results in Fig. 3 are obtained by means of the finite-element-based commercial software package COMSOL Multiphyics v. 3.4 \cite{Comsol}. In this case, we consider a computational domain composed of a single unit-cell, terminated with periodic boundary conditions along the $x$-direction, and of finite size ($L=10$ $\mu$m, with vacuum regions of $0.2$ $\mu$m at both sides) and scattering boundary conditions at the entrance and the exit facets (orthogonal to the $z$-axis) of the domain.
Moreover, we assume a TM plane-wave illumination (with $y$-directed magnetic field) normally impinging along the positive $z$-direction.
We apply an adaptive discretization of the computational domain with a maximum mesh size of $0.1 \lambda$ (which results into about $27000$ degrees of freedom), and utilize the UMFPACK direct solver, with default parameters.
The field distribution in Fig. 2(c) is obtained by averaging the $y$-directed magnetic field over the unit cell. The spatial spectrum in Fig. 2(d) is subsequently obtained via a $512$-point fast-Fourier-transform.

\textbf{Funding}. U.S. Army International Technology Center Atlantic (Grant No. W911NF-14-1-0315).


%
\end{document}